\documentclass{iau}
\usepackage{graphicx}

\title[Variable stars in NGC4833] %% give here short title %%
{Wide-field variability survey of the globular cluster NGC 4833}

\author[Grzegorz Kopacki]   %% give here short author list %%
{Grzegorz Kopacki}

\affiliation{Instytut Astronomiczny, Uniwersytet Wroc\l{}awski,\\
 Kopernika 11, 51-622 Wroc\l{}aw,  Poland \\
 email: {\tt kopacki@astro.uni.wroc.pl}}

\pubyear{2013}
\volume{301}  %% insert here IAU Symposium No.
\pagerange{???--???}
\jname{Precision Astroseismology}
\editors{W. Chaplin, J. Guzik, G. Handler, A. Pigulski, eds.}
\begin{document}

\maketitle

\begin{abstract}
We present preliminary results of the variability survey in the field of the globular
cluster NGC 4833. We observed all 34 variable stars known in the cluster. In addition, 
we have found two new SX Phoenicis stars, one new RR Lyrae star, twelve new eclipsing 
systems mostly of the W Ursae Majoris type, nine new variable red giants, and
ten new field-stars showing irregular variations. Properties of RR Lyrae stars
indicate that NGC 4833 is an Oosterhoff's type II globular cluster.

\keywords{Galaxy: globular clusters: individual (NGC4833),  stars: Population II, 
stars: variables: RR Lyrae, stars: variables: SX Phoenicis}
\end{abstract}

\firstsection % if your document starts with a section,

\section{Introduction}

Continuing our ongoing project aimed at the search and analysis of pulsating stars in globular 
clusters (see \cite[Kopacki 2013]{Kopacki13}) we present preliminary results for NGC 4833. We used image 
subtraction method (ISM, \cite[Alard \&\ Lupton 1998]{Alard98}) which works well in crowded stellar fields 
like a cluster core and thus enables detection of many variable stars, such as RR 
Lyrae and SX Phoenicis stars.

NGC 4833 is the southern globular cluster of intermediate metallicity ([Fe/H]${}=-$1.85). 
The most recent version of the Catalogue of Variable Stars in Globular Clusters 
(CVSGC, \cite[Clement et al.\ 2001]{Clement01}) listed 34 objects in the field of this cluster including 
six SX Phoenicis stars and 20 RR Lyrae stars.

\section{Observations}

We used CCD observations obtained during one-month observing run in Feb/Apr, 
2008 using 40-inch telescope at Siding Spring Observatory, Australia. They consisted of 740 
$V$-filter and 220 $I_{\rm C}$-filter CCD frames.

\section{Results}

We confirmed all variable stars found recently in the cluster core 
by \cite[Darragh \&\ Murphy (2012)]{Darragh12}. In addition, we have detected two new 
SX Phoenicis stars, one new RR Lyrae star, twelve new eclipsing 
systems mostly of the W Ursae Majoris type, nine new variable 
red giants at the tip of the RGB, and ten field-stars showing irregular 
variations. Equatorial coordinates of periodic variable stars we
observed, together with derived periods, are given in Table \ref{tab1}. New variable
stars are indicated with designations starting with letter 'n'. 

The mean period of RRab stars 
in NGC 4833 is equal to $\langle P_{\rm ab}\rangle ={}$0.701 d, and relative percentage 
of RRc stars amounts to $N_{\rm c}/(N_{\rm ab}+N_{\rm c})={}$48 \%. With these values we 
find that NGC 4833 belongs to the Oosterhoff's II group of globular 
clusters. 

Almost all observed SX Phoenicis stars show multiperiodic 
light changes (see Table \ref{tab1}) with one star, v31, exhibiting oscillation in two 
first radial modes. Moreover, we found in an RRc star v20 two 
closely-spaced frequencies.

\vskip3pt\noindent
{\em Acknowledgements}. This work was supported by the NCN grant 
no.\ 2011/03/B/ST9/02667.

\begin{table}
 \begin{center}
  \caption{Equatorial coordinates and periods of periodic stars in NGC 4833.}
  \label{tab1}
  {\scriptsize
    \begin{tabular}{|l|l|c|c|l|}\hline 
{\bf Var} & {\bf Type}& {\bf $\alpha_{2000}$} [$^{\rm h}$ $^{\rm m}$ $^{\rm s}$] & 
{\bf $\delta_{2000}$} [$^\circ$ $^\prime$ $^{\prime\prime}$] & {\bf $P$} [d]\\ 
 \hline
 v27& SXPhe& 12 59 13.73& $-$70 52 09.5& 0.0509813\\              
 v29& SXPhe& 12 59 35.31& $-$70 52 40.9& 0.059780\\              
 v30& SXPhe& 12 59 42.56& $-$70 53 04.0& 0.044256 0.045757\\
 v31& SXPhe& 12 59 47.98& $-$70 52 51.5& 0.0533323 0.041726\\             
 v32& SXPhe& 12 59 55.05& $-$70 52 24.2& 0.0706795 0.072537\\             
 v33& SXPhe& 12 59 57.64& $-$70 54 29.5& 0.0719219 0.079772\\             
 n35& SXPhe& 12 58 59.01& $-$70 52 05.1& 0.037510 0.036708\\
 n36& SXPhe& 12 59 32.64& $-$70 53 15.9& 0.040149\\             
 \hline
 v01& RRLyr& 12 58 36.53& $-$70 44 48.1& 0.750082\\              
 v03& RRLyr& 12 59 33.73& $-$70 52 13.2& 0.74453\\              
 v04& RRLyr& 12 59 33.76& $-$70 51 57.8& 0.65577\\              
 v05& RRLyr& 13 00 01.28& $-$70 53 17.3& 0.629424\\              
 v06& RRLyr& 12 59 56.18& $-$70 50 10.2& 0.65400\\              
 v07& RRLyr& 12 59 48.80& $-$70 52 21.1& 0.66888\\              
 v12& RRLyr& 12 59 37.91& $-$70 52 14.7& 0.58980\\             
 v13& RRLyr& 13 00 29.86& $-$70 52 55.9& 0.36788\\              
 v14& RRLyr& 12 59 31.04& $-$70 53 06.9& 0.40842\\              
 v15& RRLyr& 12 59 20.15& $-$70 53 25.0& 0.66745\\              
 v17& RRLyr& 12 59 43.98& $-$70 54 24.9& 0.390263\\              
 v18& RRLyr& 12 59 28.26& $-$70 54 25.5& 0.42559\\              
 v19& RRLyr& 12 59 06.02& $-$70 53 29.7& 0.370658\\              
 v20& RRLyr& 12 59 08.21& $-$70 52 23.5& 0.2997 0.3012\\             
 v21& RRLyr& 12 59 50.94& $-$70 50 36.5& 0.39878\\
 v22& RRLyr& 12 59 45.10& $-$70 53 55.5& 0.85095\\              
 v23& RRLyr& 12 59 44.74& $-$70 51 27.0& 0.406503\\              
 v24& RRLyr& 12 59 36.68& $-$70 52 58.7& 0.62612\\              
 v26& RRLyr& 12 59 02.74& $-$70 52 51.9& 0.31788\\              
 v28& RRLyr& 12 59 21.22& $-$70 53 26.4& 0.87401\\            
 n37& RRLyr& 12 59 43.16& $-$70 53 36.5& 0.30215\\              
 \hline
 v25& Ecl&   12 58 55.37& $-$70 51 45.9& 0.72310\\              
 v34& Ecl&   13 00 25.17& $-$70 49 16.4& 0.36290\\              
 n38& Ecl&   13 00 11.88& $-$70 51 26.4& 0.257129\\              
 n39& Ecl&   13 01 33.10& $-$70 49 22.6& 0.28005\\              
 n40& Ecl&   13 01 49.41& $-$70 51 42.1& 0.282581\\              
 n41& Ecl&   13 01 29.29& $-$70 44 58.0& 0.29917\\              
 n42& Ecl&   12 57 17.71& $-$70 47 51.0& 0.3137\\              
 n43& Ecl&   13 00 22.56& $-$70 55 25.8& 0.3175\\              
 n44& Ecl&   12 59 52.75& $-$70 44 32.7& 0.3362\\
 n45& Ecl&   13 00 43.49& $-$70 47 26.9& 0.362332\\              
 n46& Ecl&   13 00 24.25& $-$70 56 01.4& 0.38071\\              
 n47& Ecl&   12 59 37.69& $-$70 51 28.4& 0.48875\\             
 n48& Ecl&   13 00 45.49& $-$70 47 38.2& 0.49226\\              
 n49& Ecl&   13 00 55.03& $-$70 54 15.9& 1.1235\\              
 \hline
 n59& Ukn&   13 00 27.52& $-$70 54 01.6& 2.889\\              
 n60& Ukn&   12 59 14.62& $-$70 50 37.8& 4.390\\              
 n61& Ukn&   13 00 20.85& $-$70 49 56.4& 6.300\\             
 \hline
  \end{tabular}
  }
 \end{center}
\end{table}

\end{document}